\documentclass[twocolumn, prb, showpacs, superscriptaddress]{revtex4}

\usepackage{graphicx}
\usepackage{dcolumn}
\usepackage{bm}
\usepackage{color}

\begin{document}

\title{Raman study of coupled electronic and phononic excitations in $LuB_{12}$}

\pacs{62.50.+p, 78.30.Er, 71.18.+y, 72.10.Di}

\author{Yu. S. Ponosov}
\affiliation{M.N. Mikheev Institute of Metal Physics UB RAS, 620990, S. Kovalevskaya str. 18, Ekaterinburg, Russia}
\affiliation{Ural Federal University, Mira St. 19, 620002 Ekaterinburg, Russia}
\author{A. A. Makhnev}
\affiliation{M.N. Mikheev Institute of Metal Physics UB RAS, 620990, S. Kovalevskaya str. 18, Ekaterinburg, Russia}
\author{S. V. Streltsov}
\affiliation{M.N. Mikheev Institute of Metal Physics UB RAS, 620990, S. Kovalevskaya str. 18, Ekaterinburg, Russia}
\affiliation{Ural Federal University, Mira St. 19, 620002 Ekaterinburg, Russia}
\author{V. B. Filipov}
\affiliation{I N Frantsevich Institute for Problems of Materials Science of National Academy of Sciences of Ukraine,
3 Krzhizhanovskogo Street, 03680 Kiev, Ukraine}
\author{N. Yu. Shitsevalova}
\affiliation{I N Frantsevich Institute for Problems of Materials Science of National Academy of Sciences of Ukraine,
3 Krzhizhanovskogo Street, 03680 Kiev, Ukraine}
\begin{abstract}

Electronic Raman scattering and optical phonon self-energies are studied on single crystals of $LuB_{12}$ with different isotopic composition in the temperature region 10-650K and at pressures up to 10 GPa. The shape and energy position of the spectral peaks depend on the magnitude of the probed wave vector, temperature, and symmetry of excitations. We simulated experimental spectra using electronic structure obtained in the density function theory and taking into account the electron-phonon scattering. The emergence of a broad continuum in the spectra is identified with the inelastic scattering of light from the electronic intraband excitations. Their coupling to non-fully symmetric phonon modes is the source of both the Fano interference and temperature-dependent phonon self-energies. In addition, long wavelength vibrations of the boron atoms are in nonadiabatic regime, so the electronic contribution to their self-energies provides a temperature dependence that is similar to the anharmonic contribution. Comparison of calculation and experiment allowed us to determine the coupling constant $\lambda_{ep}$=0.32, which gives correct critical temperature of the transition to the superconducting state.
\end{abstract}

\maketitle

\section{Introduction}

A number of superconducting substances belong to different families of binary boron compounds. Among them a record superconducting transition temperature was found in $MgB_2$ ($T_c$=39K)~\cite{MgB}, hexaboride  $YB_6$ has $T_c\leq$8.4 K~\cite{YB}  and dodecaboride $ZrB_{12}$ - 6K~\cite{mat}. All of them are discussed in the framework of the conventional electron-phonon mechanism of superconductivity. The leading interaction  in  $MgB_2$ assumes strong coupling of the $E_{2g}$ optical phonon (boron atoms vibration) with electrons. High values of the electron-phonon coupling constant $\lambda$ and $T_c$ in $YB_6$ and  $ZrB_{12}$ are considered to be caused by acoustic vibrations of loosely-coupled metal atoms in boron cage~\cite{lor,lorz,tis1,tis2}. For $YB_6$ it is confirmed by calculations of the electron and phonon spectra, and electron-phonon interaction by linear response method~\cite{xu} . However, a detailed comparison of calculation and experiment has not yet been done because of the lack of neutron data on the frequencies and widths of the acoustic phonons. Such calculations have not been performed yet for $ZrB_{12}$ . 

$LuB_{12}$ is a structural analogue of $ZrB_{12}$  with a considerably lower $T_c$=0.42~\cite{Lu}. Both crystallize in the $UB_{12}$ structure (space group Fm-3m), where the metal atom is located in the center of a truncated boron octahedron $B_{24}$. Their electronic densities of states at the Fermi level are comparable. Analysis of the data on heat capacity, thermal expansion, resistivity~\cite{lorz} and optical spectroscopy~\cite{tis1,tis2} suggests that this difference in the $T_c$ is due to a stronger coupling of acoustic phonons (vibrations of the zirconium atoms) with electrons. 

 Contribution of high-frequency vibrations of the boron atoms to electron-phonon coupling has been studied theoretically for $YB_6$~\cite{xu}. It has been shown that it is small, but the comparison of calculation and experiment is missing. The phonon spectra in the region of high frequencies at normal conditions in different dodecaborides were demonstrated to be similar~\cite{rub,nem,al,werh,wer1}, but temperature and pressure effects on the dynamics and electron-phonon interaction have not been studied yet. Partly this gap can be bridged by a study of their influence on Raman active phonons. Asymmetry of the phonon lines observed in a number of borides clearly indicates their interaction with the electronic continuum. Electronic excitations are indeed observed in inelastic light scattering of hexaborides $LaB_6$ and $YB_6$~\cite{lab,yb}. Their detailed study allowed us to obtain information about the velocity of electrons and their relaxation rate which is closely related with the magnitude of the electron-phonon coupling. If  the difference in  $T_c$ between  $ZrB_{12}$ and $LuB_{12}$ is indeed due to interaction with low-frequency phonons the latter compound may serve as a reference object for study of electron-induced effects on optical vibrations.

The aim of our work is to study the effect of electron-phonon interaction on optical phonons in $LuB_{12}$. For this we performed measurements of the frequencies and widths of optical phonons as well as the spectra of electronic excitations in the wide range of temperatures (10-650К) and pressures (0-10 GPA). Calculation of the phonon self energies and spectra of electronic excitations was performed on the basis of density functional theory (DFT). Account of the carrier scattering by
phonons allowed to reproduce the anomalous temperature behavior of the self-energy of the E and T phonons, which is partly determined by the nonadiabatic interaction with the intraband electronic continuum.

\section{Experiment}
The single-crystal samples $LuB_{12}$ were grown by vertical crucible-free inductive floating zone melting in an inert
gas atmosphere in  IPM NASU (Kiev). The details  are presented in ~\cite{sam}. Measurements were carried out on samples with natural isotopic composition and samples enriched with isotopes $B^{10}$ and $B^{11}$. For measurements under pressure thin nonoriented fragments of crystals with sizes $\sim$100$\times$100$\times$10 $\mu$m were loaded into a gasketed diamond anvil cell (DAC) using a 4 : 1 methanol-ethanol mixture as a pressure medium. The pressure in the cell was controlled 
using the ruby luminescence method. For temperature measurements  single crystal samples with surface orientation (001) were placed into the optical microcryostat. The spectra were excited by 532 nm (2.33 eV) and 785 nm (1.55 eV) lines of solid-state lasers and 633 nm (1.96 eV) of He-Ne laser with power up to 5 mW, focused to a spot on the sample with a diameter of $\simeq$ 5-10 microns. To exclude surface effects, the measurements were carried out on freshly cleaved surfaces. The scattered light was detected by a Renishaw RM 1000 microspectrometer equipped with filters to exclude low-frequency Rayleigh scattering with a threshold of $\approx$50 $cm^{-1}$ and a thermoelectrically cooled CCD-based detection system. The spectral resolution was $\sim$ 2-3 $cm^{-1}$. A complete set of polarization geometries was used to extract Raman active  phonon representations of $A_{1g}$, $E_g$ and $T_{2g}$; XX ($A_{1g}$+$E_g$), XY - $T_{2g}$, X'Y' - 3/4$E_g$, where X,Y $\|$ [100], [010], and X',Y' $\|$ [110], [$\overline{1}10$]. 

For the calculation of the distribution of wave-vector transfers, ellipsometric measurements of
the optical constants \textit{n} and \textit{k} were performed. Since the optical properties of $LuB_{12}$ in the visible spectral range vary significantly, the spectra were corrected for the optical absorption, transmission and refraction, as well as the spectral response of the spectrometer. All figures below show Raman response $\chiχ^{''}(\omega)$ obtained after correction by $n_{B}(\omega, T) + 1$, where $n_{B}(\omega, T)$ is the the Bose-Einstein factor.
\begin{figure}[b]
\includegraphics[width=0.45\textwidth]{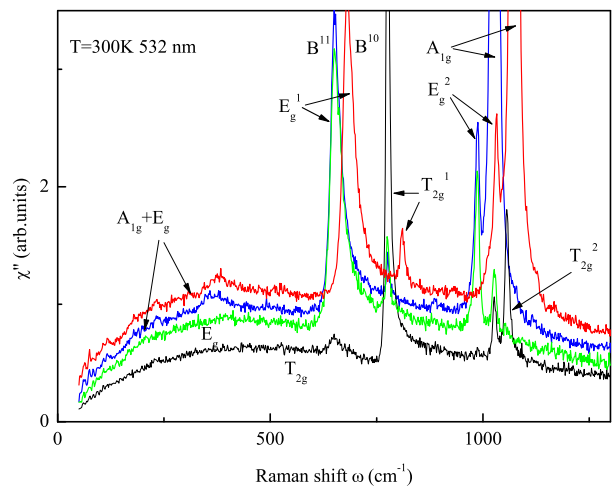}
\caption{\label{label}Raman response $\chi^{''}(\omega)$ in $LuB_{12}$, enriched by $B^{10}$ and $B^{11}$ isotopes, obtained from the (001) plane at 300K in different polarization geometries with 532 nm excitation. Intensities of three upper spectra are almost equal and the spectra were shifted for clearness.}
\end{figure}

\section{Raman results and calculations}
\subsection{Electronic response}
Fig.1 shows $LuB_{12}$ spectra measured in different polarization geometries at T=300 K on a single crystals enriched by isotopes $B^{11}$ and  $B^{11}$. They contain 5 narrow lines superimposed on a broad background, which intensity depends on the scattering symmetry. 
The continuum is strongly polarization dependent, with large $E_g$+$T_{2g}$ symmetry and negligible $A_{1g}$ symmetry contributions. The shape and position of the observed continuum do not change in measurements on the samples with different isotopic composition while the phonon lines show shifts in consistence to atomic mass ratio. Raman response $\chi^{"}(\omega)$ for different temperatures is shown in Fig.2. The spectra observed with 633 nm (1.96 eV) laser excitation   (Fig.2A) are very similar to ones measured in~\cite{slu}. The low-temperature continuum near 185 $cm^{-1}$ in these spectra shifts to higher energies and broadens with increasing temperature. As one can see the intensity of the broad continua goes to zero at all temperatures when $\omega\rightarrow$ 0. Significant broadening and shift of the peaks to $\simeq$800~$cm^{-1}$ was found with increase in temperature to 650K. When using  532 nm (2.33 eV) laser excitation the maximum of low-temperature continuum is observed near $300~cm^{-1}$ (Fig.2b). Its energy is close to the energy of the continuum at 300K, measured with 633 nm  laser excitation. As was shown in~\cite{lab,yb}, such scattering in borides originates from intraband electronic transitions near the Fermi level. The spectra independence from isotope composition confirms their relation with electronic excitations.
\begin{figure}[t]
\includegraphics[width=0.4\textwidth]{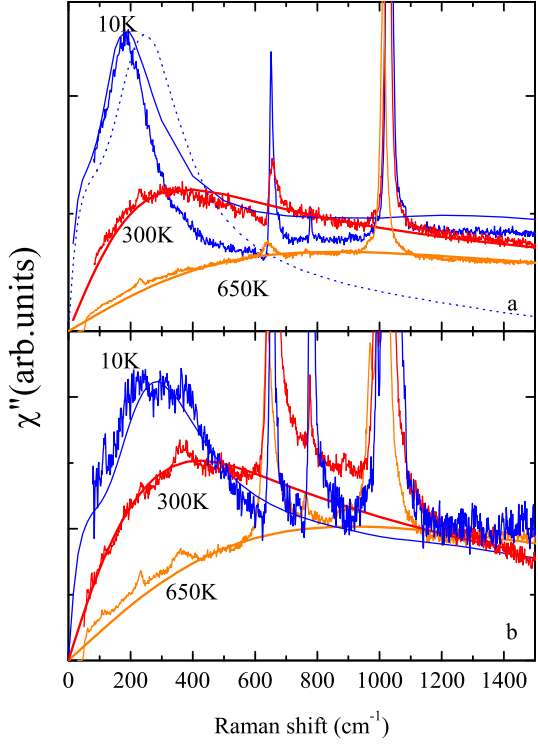}
\caption{\label{label} Raman response $\chi^{''}(\omega)$ in $LuB^{11}_{12}$, obtained from the (001) plane at various temperatures in the polarization geometry (XX) - $A_{1g}$+$E_g$ of symmetry; (a) - 633 nm excitation , (b)  532 nm excitation . The solid line represents the $\chi^{''}(\omega)$ calculated with $\lambda=0.32$. The dotted line shows the response calculated with $\lambda\approx0$.}
\end{figure}

\subsection{Optical phonons}
 The number of Raman-active phonon modes for idealized structure of the dodecaborides (space group Fm-3m) is determined by  a factor group analysis :
\begin{equation}
\Gamma_{q=0}= A_{1g} + 2E_g + 2T_{2g}.
\end{equation}
$A_{1g}$, $E_g$ and $T_{2g}$ are irreducible representations for $O_h$ point group. All five phonons are vibrations of the
boron sublattice only. The shape of the $E_g^1$ and $T_{2g}^1$ phonon lines at 650 and 790 $cm^{-1}$ (Figs. 1,2) is a striking evidence of electronic origin of broad continua. They show a strong asymmetry depending on the excitation wavelength (Fig.3) that suggests interference between phonon line and continuum. In this case the phonon lineshape may be described by the asymmetric Breit-Wigner-Fano (BWF) profile expression~\cite{klein}:
\begin{eqnarray}
I(\omega)= \pi\rho(\omega)T_e^2\frac{(Q+\epsilon)^2}{1+\epsilon^2}, \\
      \epsilon=\frac{\omega-\omega_0-V^2R(\omega)}{\Gamma},  \\       
      Q=\frac{V(T_p/T_e+V^2R(\omega)}{\pi V^2\rho(\omega)},
\end{eqnarray}
where $\omega_0$ and $R(\omega)$ are a bare (uncoupled) mode frequency and the Hilbert transform of continuum density of states $\rho(\omega)$. $T_e$ and $T_p$ are scattering amplitudes for continuum and phonon. Electron-phonon interaction with matrix element V determines the phonon width $\Gamma=2(\Gamma_0+\pi V^2\rho(\omega))$ ($\Gamma_0$ is the phonon width in absence of interaction). This interaction also shifts the phonon energy to $\omega=\omega_0+V^2R(\omega)$. A distinguishing feature of the Fano resonance in the Raman spectra is the dependence of asymmetry parameter \textit{Q} and, consequently, of the profile of the phonon line on the wavelength of the exciting light. 
\begin{figure}[b]
\includegraphics[width=0.45\textwidth]{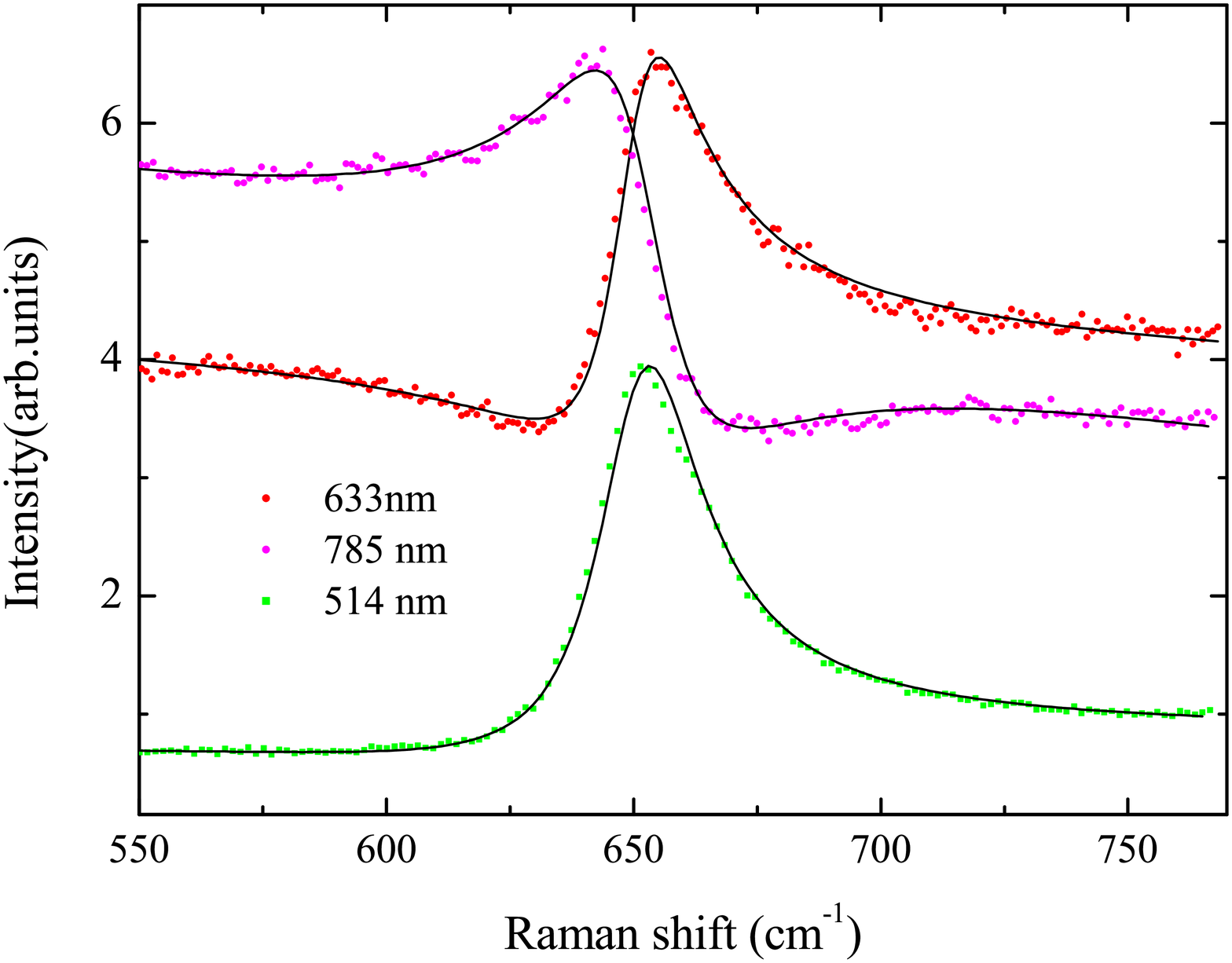}
\caption{\label{label}The lineshapes of $E_g$ phonon in $LuB_{12}^{11}$ measured with different excitation wavelengths at 300K. The solid lines show fit curves using Eq.2.}
\end{figure}
It is seen very clearly in Fig.3 where all $E_g^1$ line profiles were fitted by Equation (2). Inspite of rather large visual difference in lineshapes the extracted phonon frequencies and widths are very similar for different excitation wavelength used:  $\omega$=649.6 $cm^{-1}$, $\Gamma$=24 $cm^{-1}$ and Q=4.7 for 532 nm excitation,  $\omega$=649.2 $cm^{-1}$, $\Gamma$=23 $cm^{-1}$ and Q=2 for 633 nm excitation and   $\omega$=649.8 $cm^{-1}$, $\Gamma$=25 $cm^{-1}$ and Q=-1.4 for 785 nm excitation. Obviously, the asymmetry parameter Q changes its sign when the excitation wavelength  varies in the interval 633-785 nm. That may indicate one of the scattering amplitudes $T_e$ or $T_p$ changes its sign.

Figures 4-8 show $\omega$, $\Gamma$ and Q for 5 modes of $LuB^{11}_{12}$ as a function of temperature. The results were obtained by peak fitting to a Fano (for $E_g^1$ and $T_{2g}^1$) or a Voigt function (for 3 high-energy modes), which is a Lorentzian
folded with a Gaussian that accounts for the spectrometer bandwidth. The frequencies and widths of all phonon modes were the same when excited at both wavelengths 633 and 532 nm.
The asymmetry parameter \textit{Q} of the $E_g^1$ and $T_{2g}^1$ modes depends on the excitation wavelength and shows minima at T$\sim$400-500K.  The frequencies of all phonons soften and their widths grow with increasing temperature.  At first glance such  behavior may come from anharmonic effects. To the lowest order in perturbation theory, an account of the cubic anharmonicity leads to well-known expressions for the phonon frequency shift and linewidth~\cite{klem}:
\begin{eqnarray}
\Delta\omega_{an}(T)=\Delta\omega_{therm}+A\left[1+2n_B(\omega/2)\right], \\
\Gamma_{0}(T)=\Gamma^i+B\left[1+2n_B(\omega/2)\right] ,\\
\Delta\omega_{therm}=\omega_0\left[exp\left(-\gamma\int_0^T\alpha(T^{'})dT^{'}\right)-1\right].
\end{eqnarray}
The bare linewidth $\Gamma^i$ includes contributions from different defects, A and B are fitting parameters related to the third-order coefficients in the expansion of the lattice potential in normal coordinates.

Based on the measured pressure dependences of the phonon frequencies (Fig.9), we estimated the contributions of thermal expansion term $\Delta\omega_{therm}$ to the phonon frequency using Eq.7. Gruneisen parameters for the $E_g^1$ ($\gamma$=1.02(4)), $T_{2g}^1$ ($\gamma$=1.15(6)) and $A_{1g}$ ($\gamma$=1.38(6)) were estimated assuming a bulk modulus $B_0$ = 232 GPa~\cite{gr}. Linear thermal expansion coefficients $\alpha$(T) from~\cite{pad,mori} were used for calculation of the frequency shift $\Delta\omega_{therm}$. The low intensity of the $E_g^2$ and $T_{2g}^2$ phonons did  not allow to observe them at high pressures, so their $\gamma$ were taken to be equal to $\gamma$ for the $A_{1g}$ mode.  The obtained $\Delta\omega_{therm}$ are plotted by dotted lines for all phonon modes in Fig.4-8.
\begin{figure}[t]
\includegraphics[width=0.45\textwidth]{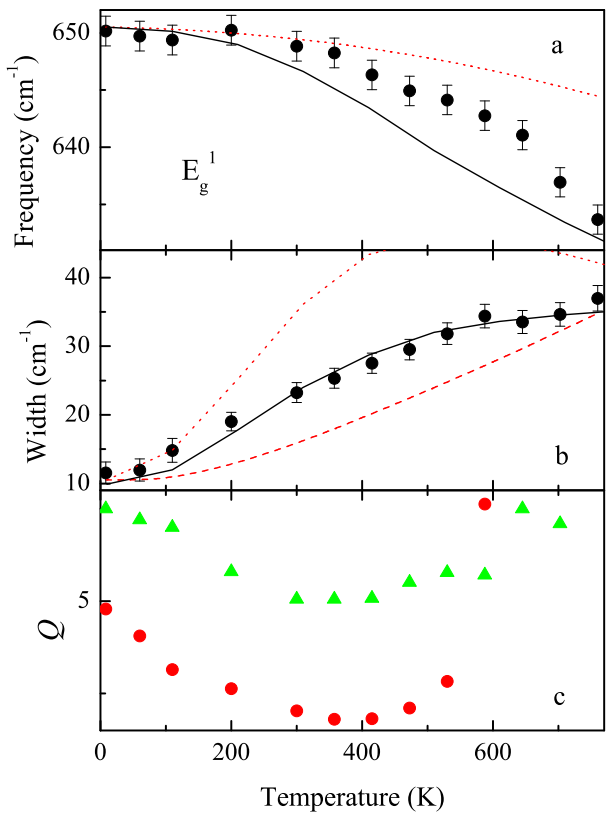}
\caption{\label{label} Temperature dependences of the frequency (a), width (b) and asymmetry parameter \textit{Q} (c) for $E_{g}^1$ phonon at 650 $cm^{-1}$. Solid line in (a) and (b)-calculated temperature dependences. The contribution of thermal expansion to the phonon frequency is shown by dotted line in (a). Width dependences are shown in (b) in the cases when the low-temperature phonon damping is fully determined anharmonicity (dashed line) and electron-phonon interaction (dotted line).}
\end{figure}
Assuming that the low-temperature phonon widths are determined solely by anharmonic processes we calculated the temperature dependence of widths for all phonons with $\Gamma^i$=0.5 $cm^{-1}$ (Eq.6).  We got excellent fit for the width of the $A_{1g}$ mode (Fig.8(b)), but poor agreement for all other modes. If the phonon linewidth is governed by decay into pairs of phonons with opposite wave vectors the linear increase of the linewidth at high temperatures should be observed. This is contrast to the experiment where high-temperature humps are observed. Varying the anharmonic coupling parameter A the anharmonic contribution to the temperature dependence of the $A_{1g}$ frequency has been calculated using Eq.5. It is plotted in Fig.8(a) showing best agreement for $\omega_0$=1042 $cm^{-1}$, A=11 $cm^{-1}$ and B=9.5  $cm^{-1}$. Fitting parameters for all phonon modes are presented in Table 1. Thus, the temperature behavior of the $A_{1g}$ phonon self-energies may be well explained by three-phonon coupling approximation. Inability to describe the damping dependences for the E and T  phonons indicates the existence of additional decay mechanisms. These may be an electron-phonon interaction or higher-order anharmonic processes. The latter provide a quadratic dependence of the phonon linewidth vs. temperature. The first one is really manifested itself through the abovementioned Fano interference, which leads to the noticeable asymmetry of the $E_g^1$ and $T_{2g}^1$ phonon profiles. Lineshape asymmetry is lesser for the $E_g^2$and $T_{2g}^2$ modes and absent for the $A_{1g}$ phonon. Obviously, this is due to the presence of the electronic continua with the  $E_g$ and $T_{2g}$ symmetry and  the absence of electronic excitations with  $A_{1g}$ symmetry. 
\begin{table}[t]
\caption{\label{tab:example}Fitting parameters used in calculations of the phonon self-energies of $LuB_{12}$.}
\begin{ruledtabular}
\begin{tabular}{lllllll}
 & $\omega_0$$(cm^{-1})$&g$(cm^{-1})$&$\Gamma^i$$(cm^{-1})$&A$(cm^{-1})$&B$(cm^{-1})$&$\lambda_{mode}$\\
$E_g^1$    & 654 &413&0.5&0&4&0.0323\\
$T_{2g}^1$ & 785 &250 &0.5&-5&2.4&0.0099\\
$E_g^2$    & 1007&290 &0.5&-13 &4&0.0103\\
$T_{2g}^2$ & 1080&270&0.5&-16&5&0.0083\\
$A_{1g}$   & 1042&  0&0.5&-11&9.5&0\\
\end{tabular}
\end{ruledtabular}
\end {table}

\subsection{Calculations}
 In order to theoretically estimate the effects of electron-phonon interaction on the phonon energies and line widths, calculations of a spectrum of electronic excitations and phonon spectral function were carried out. The phonon spectral function $I(\omega)$ was calculated by taking into account the frequency and temperature dependences of the phonon self-energies due to both the anharmonicity and electron-phonon interaction:
\begin{eqnarray}
I\left(\omega\right) = \frac{4\omega_0^{2}}{\pi}\int_{0}^{\infty}dq\times \nonumber\\
{\textstyle\frac{{U(q)\cdot\Gamma\left({q,\omega,T}
\right)}}{{\left[{\omega ^{2}-(\omega
_{0}+\Delta\omega_{an}(T))^{2}-2\omega_0 \Pi^{\prime}\left(q,\omega,T\right)} \right]^{2}+
4\omega_0^2 \Gamma^2\left(q,\omega,T\right)}}}
\end{eqnarray}
Here $\Gamma(q,\omega,T) = \Gamma_0(T) + \Pi''(q,\omega,T)$ is the line width, $\Pi'(q,\omega,T)$ and $\Pi''(q,\omega,T)$ are the real and imaginary parts of the phonon self-energy due to the electron-phonon scattering. The wave vector distribution $U(q)$ was taken to be of  skew lineshape~\cite{lou} and was estimated on the basis of experimental optical constants. The electron-induced phonon self-energy originates from intraband electronic transitions in our simulation. We used $\Pi(q,\omega,T)$ in the form~\cite{maks}:
\begin{figure}[t]
\includegraphics[width=0.4\textwidth]{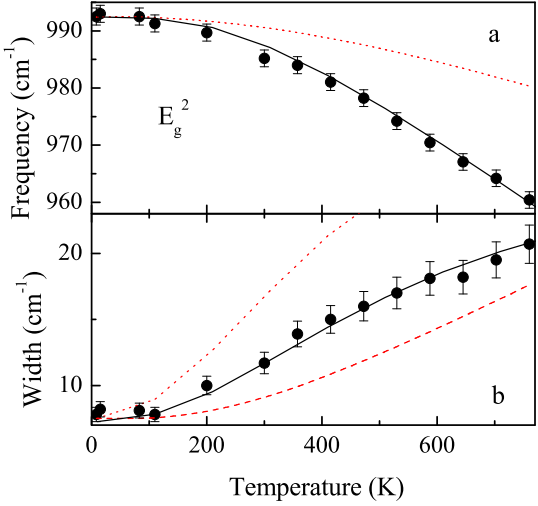}
\caption{\label{label} Temperature dependences of the frequency (a) and width (b) for the $E_{g}^2$ phonon at 990 $cm^{-1}$. Solid line in (a) and (b)-calculated temperature dependences. The contribution of thermal expansion to the phonon frequency is shown by dotted line in (a). Width dependences are shown in (b) in the cases when the low-temperature phonon damping is fully determined anharmonicity (dashed line) and electron-phonon interaction (dotted line).}
\end{figure}
\begin{eqnarray}
\Pi \left( {q,\omega,T} \right) = \oint
\frac{ds_f}{\upsilon_f } \; g^{\rm{2}} \left( {k_f,
q, \omega} \right)\; \times\nonumber\\
 \left\{ \left(  \int\limits_{ - \infty }^\infty  {d\varepsilon }
 \frac{f\left( \varepsilon  \right) - f\left(\varepsilon  + \omega  \right)}{\omega  - q \upsilon_f^z  - \Sigma
\left( \varepsilon  \right) + \Sigma \left( \varepsilon  + \omega \right)}\right)- {\rm{2}} \right\}.
\end{eqnarray}
Here, $ds_f$ is the area element of the Fermi surface, $\upsilon_f^z$ - the electron velocity, g - the matrix element of the electron-phonon interaction, $f(\varepsilon)$ is the Fermi function, and $z$ denotes the normal to the sample surface. The retarded and advanced quasi-particle electronic self-energies $\Sigma(\epsilon)$ and $\Sigma(\epsilon+\omega)$ determine the electron spectrum renormalization near the Fermi level due to different interactions. In the case of electron-phonon scattering, the real and imaginary parts are:~\cite{su}
\begin{figure}[t]
\includegraphics[width=0.4\textwidth]{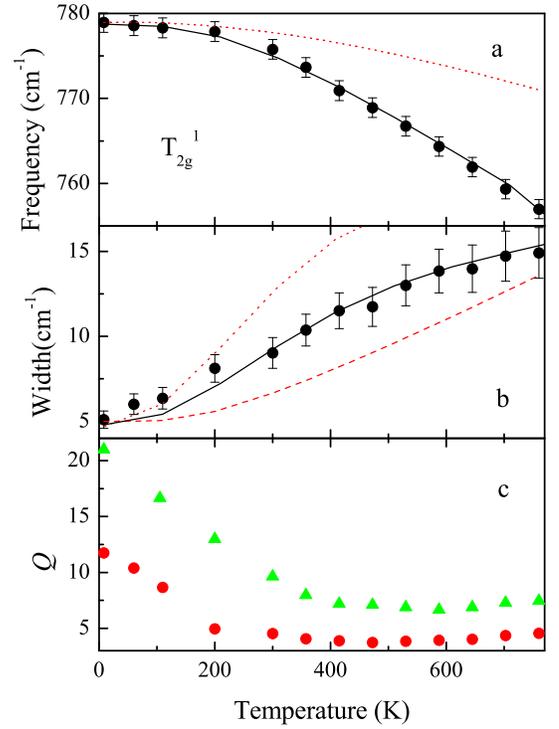}
\caption{\label{label} Temperature dependences of the frequency (a), width (b) and asymmetry parameter \textit{Q} (c) for  $T_{2g}^1$ phonon at 780 $cm^{-1}$. Solid line in (a) and (b)-calculated temperature dependences. The contribution of thermal expansion to the phonon frequency is shown by dotted line in (a). Width dependences are shown in (b) in the cases when the low-temperature phonon damping is fully determined anharmonicity (dashed line) and electron-phonon interaction (dotted line).}
\end{figure}
\begin{eqnarray}
\Sigma^{'}(\epsilon)= \int d\Omega\alpha^2F(\Omega){}\nonumber\\
{}\times \Re \left[ \psi \left( \frac{1}{2}+i\frac{\epsilon+\Omega}{2T}\right)-\psi\left(\frac{1}{2}+i\frac{\epsilon-\Omega}{2T} \right) \right],
\end{eqnarray}
\begin{eqnarray}
\Sigma^{''}(\epsilon)= \pi \int d\Omega\alpha^2F(\Omega){}\nonumber\\
\times \left[ 2n_B(\Omega)-f\left(\epsilon-\Omega\right)+f\left(\epsilon+\Omega\right)+1 \right]+\nu,
\end{eqnarray}
where $\nu$ is the impurity relaxation frequency, $\Psi$ - the digamma function, $\Omega$ - the phonon energy, $\alpha^2F(\Omega)$ - the Eliashberg spectral function for the electron-phonon interaction. 

Electronic structure and velocity of electrons at the Fermi surface were calculated using the linearized muffin-tin orbitals (LMTO)~\cite{an} in the approximation of local electron density (LDA) with exchange-correlation part of the proposed by von Barth and Hedin~\cite{hed}.  Integration over the Fermi surface was performed with a fine mesh of 125,000 k-points in the full Brillouin zone. The crystal structure parameters (space group $O_h^5$) were taken from Refs.~\cite{pad,mori}. Calculated electronic structure and the Fermi surface are consistent with the results of previous studies~\cite{gr,hei,jag}.
\begin{figure}[t]
\includegraphics[width=0.4\textwidth]{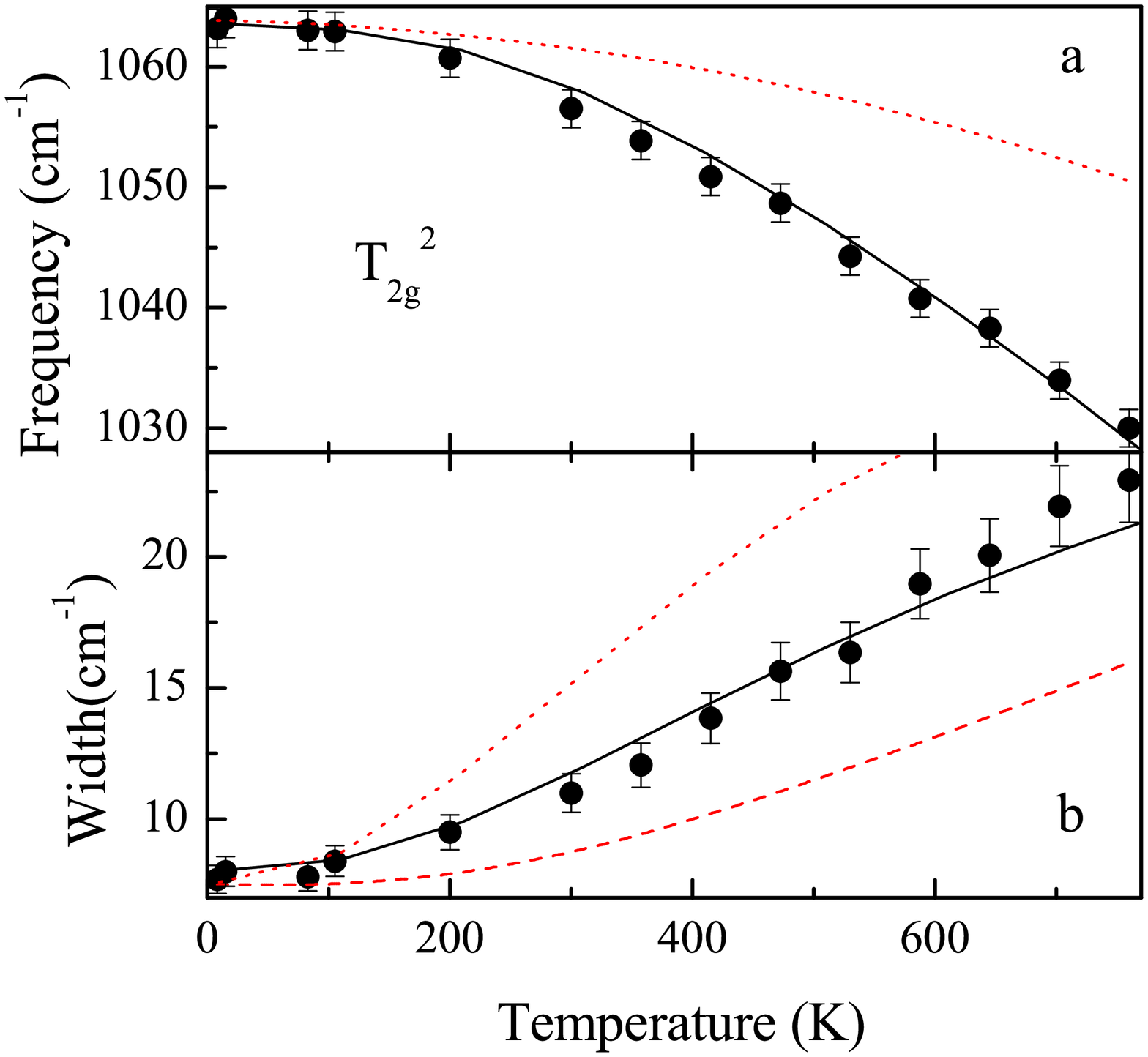}
\caption{\label{label} Temperature dependences of the frequency (a) and width (b) for the $T_{2g}^2$ phonon at 1060 $cm^{-1}$. Solid line in (a) and (b)-calculated temperature dependences. The contribution of thermal expansion to the phonon frequency is shown by dotted line in (a). Width dependences are shown in (b) in the cases when the low-temperature phonon damping is fully determined anharmonicity (dashed line) and electron-phonon interaction (dotted line). }
\end{figure}

The frequency dependence of light scattering by electronic excitations is determined by imaginary part of Eq.9 where the matrix element of electron-phonon interaction ${g}$ is replaced by matrix element of electron-photon interaction~\cite{carip}. The latter was taken by constant.  The single adjustable parameter in the calculation is the constant of electron-phonon coupling $\lambda$ = 2$\int\alpha^2F(\Omega)/\Omega$. Its starting value was determined from the evaluation of the electron-phonon scattering rate $\approx 2\Sigma^{''}(\epsilon)$ at high temperatures. To calculate $\lambda$ we used a constant $\alpha^2(\omega)$ and the phonon density of states $F(\Omega)$ from Ref.~\cite{rub}, the value of relaxation rate due to impurity scattering was taken as small as $\nu$=10 $cm^{-1}$.
\begin{figure}[b]
\includegraphics[width=0.4\textwidth]{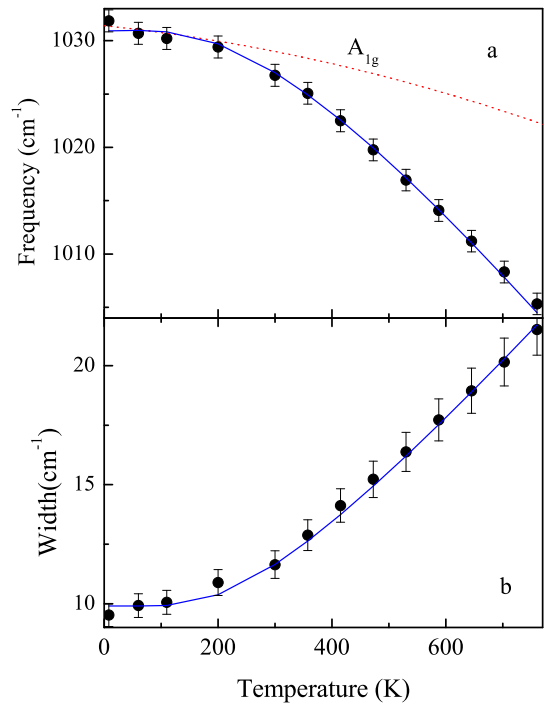}
\caption{\label{label} Temperature dependences of the frequency (a) and width (b) for the $A_{1g}$ phonon at 1030 $cm^{-1}$.The contribution of thermal expansion to the phonon frequency is shown by dotted line in (a).}
\end{figure}
As can be seen in Fig.2, the measured Raman response by electronic excitations is well described by the calculated curves at different temperatures using $\lambda$ =0.32. It follows from eq. (9)  that at low temperatures the maximum of the continuum should be observed at the frequency $\omega \simeq q\upsilon_f^z+2\Sigma^{'}(\epsilon)$, if $\Sigma^{''}(\epsilon)$ is small. Because low-frequency phonons are frozen, it is really small at T=10 K at low energies. This basically ensures collisionless regime for electrons and gives an opportunity to observe a rather narrow peak at 185 $cm^{-1}$. The conservation of the wave vector in this process allows to estimate the renormalized velocity of the electrons at the Fermi surface $\upsilon_f\simeq7.3\times 10^7$ cm/sec. An increase of the electron damping with increasing temperature leads to the appearance of incoherent scattering at high energies. This results in frequency hardening and broadening of the electronic continua. 

Variation of excitation wavelength changes the value of the probed wave vector (0.5$\times10^6$~$cm^{-1}$ for 633 nm and 0.75$\times10^6$~$cm^{-1}$ for 532 nm). Naturally, this leads to a change of a term $q\upsilon_f^z$ in (9) and results in a shift of the energy of the low-temperature electronic continuum for excitation at 532 nm.  This is in agreement with the calculation of the spectra for excitation energy of 2.33 eV in Fig.2b. Using the Allen-Dines expression for the superconducting transition temperature~\cite{all} we calculated $T_c$ for $LuB_{12}$. If Coulomb pseudopotential is taken to be $\mu^*=0.1$ the obtained value of $\lambda \approx$ 0.32 gives $T_c\approx$ 0.5 K, which is consistent with the experimental value~\cite{Lu}.

Assuming that the low-temperature widths of the E and T phonon lines are completely defined by the electron-phonon interaction we calculated phonon spectral functions according to Eq. 8 varying \textit{g} and setting A and B equal zero. The phonon profiles were fitted by Lorenzians and the temperature dependences of widths were plotted (dotted lines in Figs.4-7). They overestimate experimental widths and show maxima at intermediate temperatures. Thus, neither anharmonic mechanism nor electron-phonon mechanism is unable separately to explain the observed temperature behavior of the phonon damping. Therefore, both mechanisms anharmonicity and electron-phonon interaction give contributions to line widths and, possibly, to frequency shifts.
\begin{figure}[t]
\includegraphics[width=0.45\textwidth]{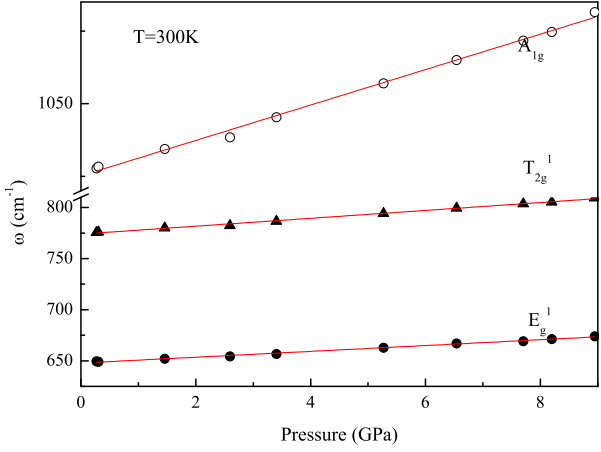}
\caption{\label{label} Pressure dependences of the phonon frequencies measured in $LuB_{12}^{11}$ at 300K and their linear fits.}
\end{figure}

Then, we fitted the observed temperature dependences of the widths and frequencies of E and T modes calculating the phonon spectral functions at different temperatures. The values A,B and \textit{g} were varied in this fit, electron-phonon coupling constant was set $\lambda$= 0.32. Calculated temperature dependences are shown in Figs. 4-7. One may see nearly perfect agreement with experimental data.

\section{Discussion}
The electronic part of the Feinman diagram for interaction of the intraband electronic excitations with phonons is isomorphic with that of light scattering by the same excitations~\cite{carip}. The imaginary part  of this self-energy $\Pi''(q,\omega,T)$  determines the phonon line width  and, correspondingly, the probability of light scattering by the electronic excitations (with replacement of \textit{g} on the electron-photon vertex). Tab. 1 shows that the highest contribution of electron-phonon interaction to the line width corresponds to the $E_g^1$ phonon: $\lambda_{E_{g}^1}$=$g^2N_f/\omega_0$ for this mode is at least three times more than $\lambda$ for other modes. Since the frequency dependence of the imaginary part of the electronic self-energy (proportional to electronic light scattering intensity) is flat in the energy range of optical phonons (Fig.2) the main reason for the difference in electronic contributions to the phonon widths is the difference in matrix elements. The low-temperature contribution of electron-phonon interaction to the linewidth of the $E_g^1$ phonon exceeds the contribution of anharmonicity by $\sim$1.5 times and this difference rises to a maximum ($\sim$2.5 times) in the region of 500~K. This happens because the maximum of the electronic continuum shifted towards higher frequencies with increasing temperature. Corresponding increase of $\Pi''(q,\omega,T)$ leads to higher electronic contributions to the width of all lines. 

As one can see, the electron-phonon mechanism  provides the same temperature trend for the phonon self-energies as the anharmonic one. The reason for this behavior is clear from the following evaluation of the phase velocity for the $E_g^1$ phonon at 654 $cm^{-1}$, which upon excitation at 633 nm yields $2.5\times10^8 cm/sec$. This value is several times larger than the average velocity of the electrons $\upsilon_f\simeq7.3\times 10^7$ cm/sec. This means that $E_g^1$ and other higher frequency phonons are in the nonadiabatic regime (Fig. 10). No Landau damping, which threshold is situated at $\approx1.2\times10^6$ $cm^{-1}$, is possible for these phonons.  It is obvious that for probed wavevectors, the phonon damping is small at low temperatures, but increases with increasing temperature until a maximum and then decreases. Such behavior explains appearance of humps in temperature dependences of the phonon damping for E and T phonons. The growth of the electron-induced contribution to phonon width is the reason of the decrease of asymmetry parameter Q for the $E_g^1$ and $T_{2g}^1$ phonons with increasing temperature up to 400-500 K, see Eq. 4. Its increase at higher temperatures may be related to the changes in scattering amplitudes by electrons or phonons.
\begin{figure}[t]
\includegraphics[width=0.45\textwidth]{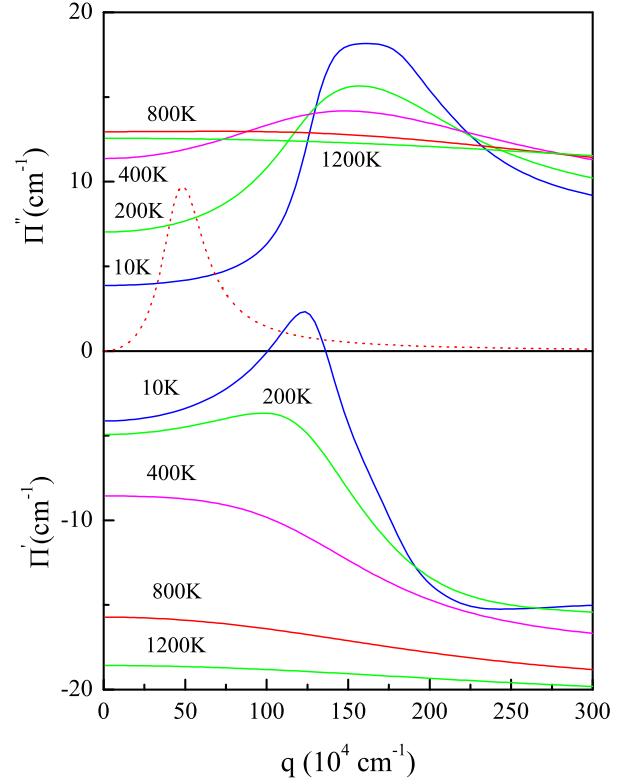}
\caption{\label{label} \textit{q} dependences of the electron-phonon self-energies  in $LuB_{12}$ for different temperatures. Probed wavevector distribution for the 633 nm excitation is shown by dotted line.}
\end{figure}

It should be noted that despite of a good description of the temperature dependence of the $E_g^1$ phonon linewidth, the shift of its frequency does not require any anharmonic contribution except the term describing thermal expansion (Tab.~1). Moreover, in the temperature region of 500-600~K the calculation overestimates the frequency softening. The ratio of electron-phonon and anharmonic contributions to the phonon widths decreases for the $E_g^2$ , $T_{2g}^1$  and $T_{2g}^2$ phonons (Tab.~1) and the anharmonic contributions to the real part of the self-energy need to be added for all of them.

The electronic light scattering as well as the self-energy of optical phonon with the full symmetry of the crystal may be screened by the carrier response in the case of  electronic bands with parabolic dispersion. This is not the case of $LuB_{12}$ which has rather anisotropic Fermi surface. However, both the intensity of the electronic light scattering of  the $A_{1g}$ symmetry and the electronic contributions to the self-energy of the $A_{1g}$ phonon are negligible.

\section{Conclusions}

Inelastic light scattering by phononic and electronic excitations in $LuB_{12}$ was studied in a wide region of temperatures and pressures for different excitation energies and scattering symmetries.  Comparison of experimental and simulated spectra based on the calculated electronic structures  was performed. It confirms that intraband electronic transitions are the source of the observed broad continua in Raman spectra. It is shown the electron-phonon scattering governs renormalization of the electron spectrum near the Fermi level. The interaction of  renormalized spectrum with non-fully symmetric  phonons is responsible for the interference effects and contributes to  phonon self-energies. The obtained value of the electron-phonon coupling constant for $LuB_{12}$ $\lambda\approx$ 0.32 gives an appropriate estimation of $T_c$ in this compound.

\section*{Acknowledgments} This work was carried out within the state assignment of the Federal Agency of Scientific Organizations of the Russian Federation (theme no. 01201463326 ``Electron'') and was supported in part by the Russian Foundation for Basic Research (project no. 14-02-00952).


\begin{thebibliography}{50}

\bibitem{MgB} J. Nagamatsu, N. Nakagawa, T. Muranaka, Y. Zenitani, J. Akimitsu,  Nature ,\textbf{41}, 63 (2001).
\bibitem{mat} B. T. Matthias, T. H. Geballe , K. Andres, E. Corenzwit , G. Hull, J. P. Maita,  Science, \textbf{159}, 530 (1968).
\bibitem{YB}Fisk, P. H. Schmidt, and L. D. Longinotti, Mat. Res. Bull. \textbf{11}, 1019 (1976).
 \bibitem{lor} ] R. Lortz, Y. Wang, U. Tutsch, S. Abe, C. Meingast, P. Popovich, W. Knafo, N. Shitsevalova, Yu.B. Paderno, and J. Junod, Phys. Rev. B 73, 024512 (2006).
\bibitem{lorz} R. Lortz, Y. Wang, S. Abe, C. Meingast, Yu. B. Paderno, V. Filippov, and A. Junod, Phys. Rev. B \textbf{72}, 024547
(2005).
\bibitem{tis1} J. Teyssier, A. B. Kuzmenko, D. van der Marel, F. Marsiglio, A. B. Liashchenko, N. Shitsevalova, and V. Filippov, Phys. Rev. B \textbf{75}, 134503 (2007).
\bibitem{tis2} J. Teyssier, R. Lortz, A. Petrovic, and D. van der Marel, V. Filippov, and N. Shitsevalova, Phys. Rev. B \textbf{78}, 134504 (2008).
\bibitem{xu} Y. Xu, L. Zhang, T. Cui, Y. Li, Y. Xie, W. Yu, Y. Ma, and G. Zou, Phys. Rev. B 76, 214103 (2007).
\bibitem{Lu} K. Flachbart, S. Gabani, K.Gloos, M. Meissner, M. Opel, Y. Paderno, V. Pavlik, P. Samuely, E. Schuberth, N. Shitsevalova, K. Siemensmeyer, and P. Szabo,  J. Low Temp. Physics, \textbf{140}, 339 (2005).
\bibitem{rub} A. V. Rybina, K. S. Nemkovski, P. A. Alekseev et al., Phys. Rev. B 82, 024302 (2010).
\bibitem{nem} K. S. Nemkovski, P. A. Alekseev, A. V. Rybina, J.-M. Mignot, K. Flachbart, P. Samuely,N. Yu. Shitsevalova, Yu. B. Paderno, F. Iga, T. Takabatake, V. N. Lazukov, E. V. Nefeodova, I. P. Sadikov, N. N. Tiden, and R. I. Bewley, Crystallography Reports,  \textbf{51},  S139 (2006).
\bibitem{al} P .A . Alekseev, Physics-Uspekhi, \textbf{58}, 330 (2015).
\bibitem{werh}H. Werheit, Yu. Paderno, V. Filippov, V. Paderno, A. Pietraszko, M. Armbruster, U. Schwarz, Journal of Solid State Chemistry, \textbf{179}, 2761  (2006).
\bibitem{wer1}  H. Werheit, V. Filipov, K Shirai, H Dekura, N. Shitsevalova, U. Schwarz, M Armbruster, J. Phys.: Condens. Matter., \textbf{23}, 065403 (2011).
\bibitem{lab}  Yu. S. Ponosov and S. V. Streltsov, JETP Lett., \textbf{97}, 447 (2013).
\bibitem{yb}  Yu. S. Ponosov, A. A. Makhnev, S. V. Streltsov,  V. B. Filipov, and N. Yu. Shitsevalova, JETP Lett. \textbf{102}, 503 (2015).
\bibitem{sam} A. Czopnik, N. Shitsevalova, V. Pluzhnikov, A. Krivchikov, Yu. Paderno, and Y. Onuki, J. Phys.: Condens. Matter, \textbf{17}, 5971 (2005).
\bibitem{slu} N. E. Sluchanko, A. N. Azarevich, A. V. Bogach, I. I. Vlasov, V. V. Glushkov, S. V. Demishev, A. A. Maksimov, I. I. Tartakovskii, E. V. Filatov, K. Flachbart, S. Gabani, V. B. Filippov, N. Yu. Shitsevalova, V. V. Moshchalkov, JETP \textbf{113}, 468 (2011).
\bibitem {klein} M. V. Klein, in Light Scattering in Solids, edited by M. Cardona  (Springer-Verlag, Berlin, 1975), p. 174.
\bibitem{klem} P. G. Klemens, Phys. Rev. \textbf{148}, 845 (1966).
\bibitem{gr} G. E. Grechnev, A. E. Baranovskiy, V. D. Fil, T. V. Ignatova, I. G. Kolobov, A. V. Logosha, N. Yu. Shitsevalova, V. B. Filippov, and O. Eriksson, Low Temp. Phys. \textbf{34}, 921 (2008).
\bibitem{pad} Yu. B. Paderno, V. V. Odintsov, I. I. Timofeeva and L. A. Klochkov, Teplofizika Vysokikh Temperatur, \textbf{9}, 200 (1971).
\bibitem{mori} T. Mori, R. Gumeniuk, Y. Grin, L. Vasylechko, Dementiy Gabunia, and N. Shitsevalova, Hasylab Annual Report, Part I, 873 (2007).
\bibitem{lou} A. \ Dervisch and R. \ Loudon, J. Phys. C {\bf 9}, L669 (1976).
\bibitem{maks} E.\ G.\ Maksimov and S.\ V.\ Shulga, Solid State Commun., {\bf 97}, 553 (1996).
\bibitem{su} S.\ V.\ Shulga, O.\ V.\ Dolgov, and E.\ G.\ Maksimov, Phys. C {\bf 178}, 266 (1991).
\bibitem{an} O.K. Andersen and O. Jepsen, Phys. Rev. Lett. \textbf{53}, 2571 (1984).
\bibitem{hed}  U. von Barth and L. Hedin, J. Phys. C\textbf{5}, 1629 (1971).
\bibitem{hei} M. Heinecke , K. Winzer , J. Noffke  al., Z. Phys. B \textbf{98}, 231 (1995).
\bibitem{jag} B. Jager, S. Paluch, O. J. Zogal et al., J. Phys.: Condens. Matter, \textbf{18}, 2525  (2006).
\bibitem{carip} M. Cardona and I. P. Ipatova, in Elementary Excitations in Solids, edited by J. L. Birman, C. Sebenne and R. F. Wallis  (Elsevier Science Publishers B.V., 1992), p. 238.
\bibitem{all}P. B. Allen and R. C. Dynes, Phys. Rev. B\textbf{ 12}, 905 (1975).

\end{thebibliography}
\end{document}